\documentclass[prd,twocolumn,notitlepage,showpacs,preprintnumbers,amsmath,amssymb,nofootinbib,APS,10pt,superscriptaddress]{revtex4-1}

\usepackage{dcolumn}
\usepackage{bm}
\usepackage{xcolor}
\usepackage[applemac]{inputenc}
\usepackage[spanish,english]{babel}
\usepackage{amsmath,amssymb,amsfonts,latexsym,cancel}
\usepackage{graphicx}
\usepackage{color}
\usepackage{soul}
\usepackage{ulem}
\usepackage{hyperref}
\usepackage{amsmath}
\usepackage{slashed}

\allowdisplaybreaks

\newcommand{\be}{\begin{equation}}
\newcommand{\ee}{\end{equation}}
\newcommand{\bea}{\begin{eqnarray}}
\newcommand{\eea}{\end{eqnarray}}

\begin{document}

\title{{\bf Translational  anomaly of chiral fermions in two dimensions}}

\author{Pau Beltr\'an-Palau}\email{pau.beltran@uv.es}
\author{Jose Navarro-Salas}\email{jnavarro@ific.uv.es}
\author{Silvia Pla}\email{silvia.pla@uv.es}

\affiliation{Departamento de Fisica Teorica and IFIC, Centro Mixto Universidad de Valencia-CSIC. Facultad de Fisica, Universidad de Valencia, Burjassot-46100, Valencia, Spain.}

\begin{abstract}

It is well known that  a quantized two-dimensional Weyl fermion  coupled to gravity spoils general covariance and breaks the  covariant conservation of the energy-momentum tensor. 
In this brief article, we point out that the quantum conservation of the momentum can also fail  in flat spacetime, provided the Weyl fermion is coupled to a time-varying homogeneous electric field. 
This signals a quantum anomaly of the space-translation symmetry, which has not been highlighted in the literature so far.\\


\end{abstract}

\date{\today}
\maketitle
\section{Introduction}\label{Introduction}

Symmetries and their corresponding Noether conservation laws play a major role in classical physics.  
It was long thought that symmetries and conservation laws are preserved  in the  quantization  of the classical system. For example, the momentum of a  classical system possessing the space-translation invariance  is a conserved quantity, and it is expected to be also conserved in the quantum theory.   In the same way, invariance under phase transformations  implies charge conservation, and it is  also expected that, after quantization, the charge operator is conserved in time. In some special situations, a classical symmetry cannot be maintained in the procedure of quantization. This happens most frequently in  field theory,  in which one encounters intrinsic  ultraviolet divergences. The removal of these infinities, through the process of renormalization, might produce finite and unambiguous results that may imply an unavoidable  conflict with the symmetry of the classical theory. 

This was first discovered in the analysis of a quantized Dirac field $\psi$ in the presence of an electromagnetic background \cite{ABJ, Jackiw}. The classical action for a massless Dirac field is invariant under chiral transformations $\psi \to e^{-i\epsilon \gamma^5}\psi$.  This implies, via  Noether's theorem, that the axial current
$j_A^\mu = \bar \psi \gamma^\mu \gamma^5 \psi$ is a conserved current $\partial_\mu j_A^\mu =0$. However, in the quantized theory this is no longer true. One finds the following  nonzero  vacuum expectation value 
\be \label{ABJ} \langle \partial_\mu j_A^\mu \rangle =   -\frac{q^2}{16\pi^2} \epsilon^{\mu\nu\alpha \beta}F_{\mu\nu} F_{\alpha \beta}\ , \ee
where $F_{\mu\nu}$ is the  electromagnetic field strength. This is a quantum breaking of the original  symmetry, and it is usually referred to as an anomaly.

Equation (\ref{ABJ}) reflects an anomaly in a global symmetry, and it allows us to better understand the underlying physics. However,  anomalies in currents coupled to gauge fields make the theory  ill defined. 
They imply an unavoidable obstruction  to  constructing the quantized theory,  and only their exact cancellation   can restore the physical consistency. 
For example, in quantum electrodynamics with a single charged Weyl fermion, we have $\langle \partial_\mu j^{\mu} \rangle \neq 0 $, and hence the theory is inconsistent. 
However, by adding a charged Weyl fermion of opposite chirality,  consistency is restored. This type of anomaly can only occur in even-dimensional spacetimes. 

A different class of gauge anomalies involves the breaking of  general covariance, reflected in the nonzero expectation values in the divergence of the energy-momentum tensor  $\langle \nabla_\mu T^{\mu\nu} \rangle \neq 0$.  They are called gravitational anomalies \cite{AGW}. These anomalies  can occur in theories with chiral fields coupled to gravity and in spacetimes of dimension $4k + 2= 2, 6, \cdots$\ , where $k$ is an integer (for a review on anomalies, see Ref. \cite{book}).

In two dimensions, one can construct very simple examples of quantum anomalies.  A Dirac field interacting with an external electromagnetic field has a chiral anomaly,
\be \langle \partial_\mu j_A^\mu \rangle =   -\frac{q}{2 \pi} \epsilon^{\mu\nu}F_{\mu\nu} \ . \ee
This implies that a (right-handed) Weyl field interacting with an external electromagnetic field possesses a harmful anomaly in the source current to which the gauge field  is coupled. 
The classical $U(1)$ local gauge symmetry is broken at the quantum level. A  chiral field  in two dimensions also possesses a  gravitational anomaly \cite {AGW, book, RW},
\be \langle \nabla_\mu T^{\mu}_{\nu} \rangle = \frac{1}{96\pi \sqrt{-g}} \epsilon^{\alpha \beta}{\partial_\beta}\partial_{\rho} \Gamma^{\rho}_{\nu \alpha} \ . \ee
It signals the breaking of the spacetime coordinate reparametrization group.

 The purpose of this paper is to point out that the  breaking of a relevant spacetime symmetry  could also happen in the quantization of a two-dimensional Weyl field,  not in the presence of gravity but in the presence  of a homogeneous electric background $E= E(t)$.  In this case, a charged  Weyl field possesses translation invariance in the spatial direction.
In the classical theory, one has conservation of the 01 component of the canonical stress-energy tensor $\partial_\mu T^{\mu1} =0$. However, in the quantized theory we find the anomalous result
\be \label{manomaly}\partial_\mu \langle T^{\mu1}_{R,L} \rangle ={\mp} \frac{q^2 A\dot A}{2\pi} \ , \ee
for right-/left-handed Weyl fields, respectively, where $A(t)$ is the vector potential for the electric field $E(t) = - \dot A(t)$. 

The result (\ref{manomaly}) has not been stressed in the previous literature, and it can be easily obtained using the method of adiabatic regularization. The adiabatic subtraction method was originally introduced to deal with ultraviolet divergences of quantized scalar fields in a homogeneous expanding universe \cite{parker-fulling, parker-toms, birrell-davies}. It has been extended to  quantized Dirac fields in the presence of a homogenous electromagnetic background in Refs. \cite{FN, BFNV, FNP}.

\section{Translational anomalies and adiabatic regularization}

Let us consider a quantized Dirac field interacting with an external homogeneous electric field $E(t)$.
The classical action for the Dirac field is given by
\bea \label{action0}
\mathcal{S}=\int d^2x \ (\frac{1}{2}\bar{\psi}i\gamma^\mu \overset{\leftrightarrow}D_\mu \psi - m\bar{\psi}\psi) \ , 
\eea
where $D_{\mu} \equiv \partial_{\mu} -i q A_{\mu}$ and   $\gamma^{\mu}$ are the Dirac matrices satisfying the anticommutation relations $\{\gamma^{\mu},\gamma^{\nu}\}=2\eta^{\mu\nu}$.
The corresponding Dirac equation  reads
 \bea
 (i \gamma^{\mu}D_{\mu}-m)\psi=0\label{diraceq} \  .
 \eea
For our purposes, it is very convenient to express the electric field in terms of a homogeneous vector potential $ E(t) = -\dot A(t)$. 
The Dirac equation \eqref{diraceq}, with $A_\mu=(0,-A(t))$, becomes (we follow here Refs. \cite{FN, BFNV}) 
  \bea
 \label{Dirac}\left(i \gamma^0 \partial_0 +\left(i\partial_x-q A\right)\gamma^1-m\right)\psi=0.
 \eea
From now on, we will use  the Weyl representation (with $\gamma^5 \equiv \gamma^0\gamma^1$)
\bea
\gamma^0 = \scriptsize
\left( {\begin{array}{cc}
 0 & 1  \\
 1& 0  \\
 \end{array} } \right),\hspace{0.5cm} 
\gamma^1 = \scriptsize \left( {\begin{array}{cc}
 0 & 1  \\
 -1& 0  \\
 \end{array} } \right), \hspace{0.5cm} \gamma^5 = \scriptsize \left( {\begin{array}{cc}
 -1 & 0  \\
 0& 1 \\
 \end{array} } \right) \nonumber
 \ . \eea
\\
We expand the field in  momentum modes 
\bea
\label{spinorbd}\psi(t, x)=\int_{-\infty}^{\infty} dk \left[B_k u_k(t, x)+D^{\dagger}_k v_k(t, x)\right] \ , 
\eea
where the two independent spinor solutions are
\bea
 u_{k}(t, x)&=&\frac{e^{ikx}}{\sqrt{2\pi }} \scriptsize \left( {\begin{array}{c}
 h^{I}_k(t)   \\
 -h^{II}_k (t) \\
 \end{array} }\right)\nonumber \\
  v_{k}(t, x)&=&\frac{e^{-ikx}}{\sqrt{2\pi }} \scriptsize \left( {\begin{array}{c}
 h^{II*}_{-k} (t)  \\
 h^{I*}_{-k}(t)  \\
 \end{array} } \right)\nonumber
 \label{spinorde}
\ . \eea
$B_k$ and $D_k$ are the creation and annihilation operators, which fulfill the usual anticommutation relations. Equation (\ref{Dirac}) is converted into 
\bea \label{system}
&&\dot{h}^{I}_k-i\left(k+qA\right)h^{I}_k-i m h^{II}_k=0\\ 
&&\dot{h}^{II}_k+i\left(k+qA\right)h^{II}_k-i m h^{I}_k=0 \label{system2}\ , 
\eea
where we assume the normalization condition $|h_k^{I}|^2+|h_k^{II}|^2=1$, ensuring the usual anticommutation relation between creation and annihilation operators. 
In the massless case, we have a decoupled system, and it can be solved analytically.

In the presence of an external homogeneous electric field,  the theory (\ref{action0}) possesses the translational invariance in the space coordinate: $x^1 \to x^1 +\epsilon$. Therefore,  Noether's theorem ensures that the  classical energy-momentum tensor $T^{\mu\nu}$ obeys the conservation law $\partial_\mu T^{\mu 1}=0$. This happens for every value of the mass. For the study of chiral conservation laws, we can use the simplest (canonical) form of the energy-momentum tensor
\be \label{tensor}
T^{\mu\nu}=\frac{i}{2}\bar{\psi} \gamma^\mu\overset{\leftrightarrow}{\partial^\nu} \psi 
\  \ee
and split it into the two chiral components $T^{\mu\nu}= T^{\mu\nu}_R + T^{\mu\nu}_L$:
\bea
&T^{\mu\nu}_R=\frac{i}{2}\bar{\psi} \gamma^\mu\overset{\leftrightarrow}{\partial^\nu} \frac{I+\gamma^5}{2}\psi \\
&T^{\mu\nu}_L=\frac{i}{2}\bar{\psi} \gamma^\mu\overset{\leftrightarrow}{\partial^\nu} \frac{I-\gamma^5}{2}\psi \ . \eea
We note that, for a massless theory and as a consequence of the underlying symmetry, the $\mu 1$ component of each chiral Weyl sector  is separately conserved   
\bea\label{DT01c}
 \partial_\mu  T^{\mu1}_{R, L} = 0 \ .
  \eea
We will show that this is no longer true in the quantum theory.  
We recall that, when $A=0$, the chiral fields $\psi_{R,L} = \frac{I{\pm} \gamma^5}{2}\psi $ obey the equations $\partial_+ \psi_R = 0 = \partial_- \psi_L$, with $x^{\pm} = x^0 \pm x^1$. The quantized fields $\psi_{R,L}$ describe particles and antiparticles traveling to the right/left, with positive/negative spatial momentum, respectively. 

The formal vacuum expectation values of these currents, for a generic value of the mass, take the form
\bea \label{CHcurrent0}
\langle T^{01}_R \rangle&=&\frac{1}{2\pi} \int_{-\infty}^{\infty} dk \,k\,|h_k^I|^2  ,\\
\langle T^{01}_L \rangle&=&\frac{1}{2\pi} \int_{-\infty}^{\infty} dk \,k\, |h_k^{II}|^2   .\
\eea
These expressions  are divergent, and we have to add appropriate subtractions. Since we are working with a homogeneous background, it is very convenient  to use the adiabatic regularization method. The method works with subtractions derived from the adiabatic expansion of the modes  \cite{parker-fulling, parker-toms} . Following Refs. \cite{FN, BFNV, FNP}, one can univocally determine the subtractions required in the renormalization of the above chiral currents.  $A(t)$ is considered of adiabatic order 1, as explained in Refs. \cite{FN, FNP}.  For an arbitrary mass, and  assuming that at early times $A(t)$ vanishes, the  renormalized expression for $\langle T^{01}_{R, L} \rangle_{ren}$ is given by 
{\small \bea \label{CHdermom}\langle T^{01}_{R,L} \rangle_{ren}=\frac{1}{2\pi }\int_{-\infty}^{\infty} dk \,k \,( |h_k^{I,II}|^2  -\frac{\omega\mp k}{2\omega}\mp\frac{3km^2 q^2 A^2}{4\omega^5})\, , \ \ \ \ \ \ 
\eea}with $\omega= \sqrt{k^2 + m^2}$.
 We can now evaluate the time derivative of the above expressions,
\bea \label{CHdermom1}
\partial_t\langle T^{01}_{R,L} \rangle_{ren}
=\pm \frac{m}{\pi} \int_{-\infty}^\infty \,k\,\mathrm{Im}(h_k^ I h_k^{II*}) dk\mp \frac{q^2 A\dot A}{2\pi} \ , \ 
\ \ \eea
where we have used the equations for the modes \eqref{system} and \eqref{system2}. 
In the massless limit, the first  term in (\ref{CHdermom1}) vanishes, and  we are left with 
\be\label{dPL}
\partial_\mu\langle T^{\mu1}_{R,L} \rangle_{ren}= {\mp} \frac{q^2 A\dot A}{2\pi} \ .
\ee
This nonvanishing result shows the existence of an anomaly in the classical translational symmetry for each chiral sector. Furthermore, this anomaly is accompanied by the well-known anomaly for the R/L  currents, 
\be \label{RLcanomalies}\partial_\mu \langle j^{\mu}_{R,L} \rangle_{ren}= {\pm} \frac{q \dot A}{2\pi} = {\mp} \frac{q}{4\pi} \epsilon^{\mu\nu} F_{\mu\nu} \ ,  \ee
where $ j^{\mu}_{R,L}= \bar \psi_{R,L} \gamma^\mu \psi_{R,L}$, which can also be derived in a similar way from the adiabatic subtractions. 
For a massless Dirac field, the anomalies cancel out, and one restores the translational invariance $\partial_\mu ( \langle T^{\mu1}_R\rangle_{ren}+ \langle T^{\mu1}_L\rangle_{ren}) =0$, as well as the phase invariance  $\partial_\mu ( \langle j^{\mu}_R\rangle_{ren}+ \langle j^{\mu}_L\rangle_{ren}) =0$.\\

\subsection{Symmetric stress-energy tensor and translational anomaly}
The  anomaly (\ref{dPL}) in the translational symmetry can also be realized in terms of the (symmetric) Belinfante stress-energy tensor $\Theta^{\mu\nu}$, constructed as 
\be \Theta^{\mu\nu} = \frac{i}{4}(\bar{\psi} \gamma^\mu\overset{\leftrightarrow}{D^\nu}\psi + \bar{\psi} \gamma^\nu\overset{\leftrightarrow}{D^\mu}\psi) \ . 
\ee
We have to remark that, although the canonical stress-energy tensor is more appropriate to show the existence of 
the translational anomaly, it is the Belinfante 
stress-energy tensor the right one to understand the 
anomaly in terms of the underlying process of particle creation.

The symmetric tensor  $\Theta^{\mu\nu}$ is related to the canonical one $T^{\mu\nu}$ by 
\be \label{theta}\Theta^{\mu\nu}= T^{\mu\nu} + \partial_\alpha B^{\alpha \mu\nu} + q\bar \psi \gamma^\mu \psi A^\nu \ , \ee
where the antisymmetric tensor $B^{\alpha\mu\nu}$ is defined as $B^{\alpha\mu\nu}= \frac{1}{8}\bar \psi \{\gamma^\alpha, \sigma^{\mu\nu} \} \psi $, and $\sigma^{\mu\nu}= \frac{i}{2}[\gamma^\mu, \gamma^\nu]$. The  divergence of the vacuum expectation values $ \langle \Theta^{\mu1}_{R,L} \rangle $ can be read from (\ref{theta})
\be \partial_\mu \langle \Theta^{\mu1}_{R,L}\rangle =  \partial_\mu \langle T^{\mu1}_{R,L}\rangle  + q(\partial_\mu A^1) \langle j^\mu _{R,L}\rangle  + qA^1\partial_\mu \langle j^{\mu}_{R, L} \rangle \ . \ee
Now, taking into account (20) and the facts that $A^1=A(t)$ and $A(t=-\infty)=0$, we obtain
\be \label{j0}\langle j^0_{R, L} \rangle_{ren} =  {\pm} \frac{qA}{2\pi}  \ . \ee
Using the result for the translational anomaly (\ref{dPL}), the anomalies for the R/L currents, and Eq. (\ref{j0}), we get immediately
\be \label{dPL2}\partial_\mu \langle \Theta^{\mu1}_{R,L} \rangle_{ren} = {\pm} \frac{q^2 A\dot A}{2\pi} = {\pm} \frac{q^2}{2\pi} E(t) \int^t_{-\infty} E(t') dt' \ ,  \ee
which can be regarded as parallel to the result (\ref{dPL}).
Note, however, the important change of sign, as compared to (\ref{dPL}). 

It is also interesting to evaluate the rate of the $00$ component of the stress-energy tensor. Using adiabatic regularization we find 
\bea \partial_\mu\langle \Theta ^{\mu0} \rangle_{ren}=  \frac{q^2A \dot A} {\pi} = -qF^{0} _{ \ 1}\langle j^1\rangle_{ren}\ . \eea
The above results can be reexpressed in null coordinates $x^{\pm}$ maintaining locality, Lorentz-covariance, and gauge-invariance.   
It is not difficult to get
\bea \label{partial1a}\partial_+ \langle \Theta_{--} \rangle_{ren} &=& -q F_{-+}\ \langle j_- \rangle_{ren}     \\
\label{partial1b} \partial_- \langle \Theta_{++} \rangle_{ren}  &=& -qF_{+-}\ \langle j_+\rangle _{ren} \ . \eea
To visualize the anomalous behavior we have to take second derivatives of the stress-energy tensor. We find
$\partial^2_+ \langle \Theta_{--} \rangle_{ren} = -q\partial_+ F_{-+}\ \langle j_- \rangle _{ren}+ \frac{q^2}{2\pi} F_{-+}^2 $ and a similar relation for $\langle \Theta_{++} \rangle_{ren}$. 
 The anomalous c-number terms in the second derivatives of the stress-energy tensor components are linked to the standard anomalous behavior of the chiral currents  $\partial_{\pm} \langle j_{\mp} \rangle_{ren} = \pm q(2\pi)^{-1} F_{+-}$. Note that $\Theta_{++}$ and $\Theta_{--}$ are related to the energy flux of the left and right moving sectors, respectively,  of the Dirac field. Note also that, in flat space, the trace of the two-dimensional stress tensor is zero $\langle \Theta_{+-} \rangle_{ren} =0$. 
 Summing up the second-order equations, one also gets
\be \label{covariantanomaly}\partial_\mu \partial_\nu \langle \Theta^{\mu\nu} \rangle_{ren} =  -q \partial_\nu F^{\nu}_{\  \rho}\langle j^\rho \rangle_{ren}  -\frac{q^2}{2\pi} F_{\mu\nu}F^{\mu\nu} \ . \ee

The quantum theory, mainly due to the above c-number terms, breaks the conservation of the chiral fluxes of momenta in a way compatible with the anomalous behavior of the chiral currents. The underlying reason for  all the above anomalies finds its origin in a  particle creation phenomenon.

\subsection{Relation to particle creation}
The  result (\ref{dPL2}) can be understood in terms of the well-known process of particle creation. Following the Bogoliubov transformation method \cite{parker-toms}, the field modes $h_k^I$ and $h_k^{II}$ for a pulsed electric field can be related, at late times, to the number density of created particles $n_k$. After some calculations, one obtains the following relations in the massless limit:

\be \label{thetabeta}
\small{\langle \Theta^{01}_{R} \rangle_{ren} =  \int_{0}^{\infty} dk \,k\, n_k \ , \ \,\,\,\,  \langle \Theta^{01}_{L} \rangle_{ren} =  \int_{-\infty}^{0} dk\,k\, n_k
\ .} \ee
It is clear that the R(L) part of the symmetric tensor gives the total momentum of the created quanta with positive (negative) momentum. Assuming $A=0$ at early times, the number density $n_k$ in the massless case is $(2\pi)^{-1}$  into the interval $(-qA(t),qA(t))$ and 0 for any other $k$ \cite{Jackiw, PSbook}. Integrating \eqref{thetabeta} between these limits, one obtains a result in full agreement with \eqref{dPL2}.

The physical picture of the underlying particle production process  is significantly modified by the mass. Let us consider a positive electric pulse $E(t) >0$. Massless   particles with positive charge are always created  with positive momentum in the interval  $(0, |qA(t)|)$, while antiparticles with negative charge are created with momentum in the interval $ (-|qA(t)|, 0)$. For massive fermions, a fraction of particles with positive charge  can be created with negative momentum, while antiparticles with negative charge can also be created with positive momentum. 

Finally, we remark that a somewhat similar result can also be obtained for each chiral sector of quantized massless scalar fields. However, the result (\ref{dPL2}) is only valid in the adiabatic limit, for an infinitely slow evolution of $A(t)$. In contrast, the result for fermions is completely general, valid for arbitrary $A(t)$. 
\\

\subsection{Relation to backreaction equations}
Another way to illustrate the  translational anomaly (\ref{dPL}) is by solving the semiclassical 
 backreaction equations for the quantized Dirac field $\psi=\psi_R+\psi_L$ 
 obeying  the Maxwell equation $ \dot E  = -q \langle j^1\rangle_{ren}$. According to the adiabatic subtraction method, we have
\be \small{\langle j^1 \rangle_{ren}=\frac{1}{2\pi} \int_{-\infty}^{\infty} dk \left( |h_k^I|^2 -  |h_k^{II}|^2  -\frac{k}{\omega}+\frac{q m^2} {\omega^3}A\right) \ .}  \ee

\begin{figure}[htbp]
\begin{center}
\begin{tabular}{c}
\includegraphics[width=90mm]{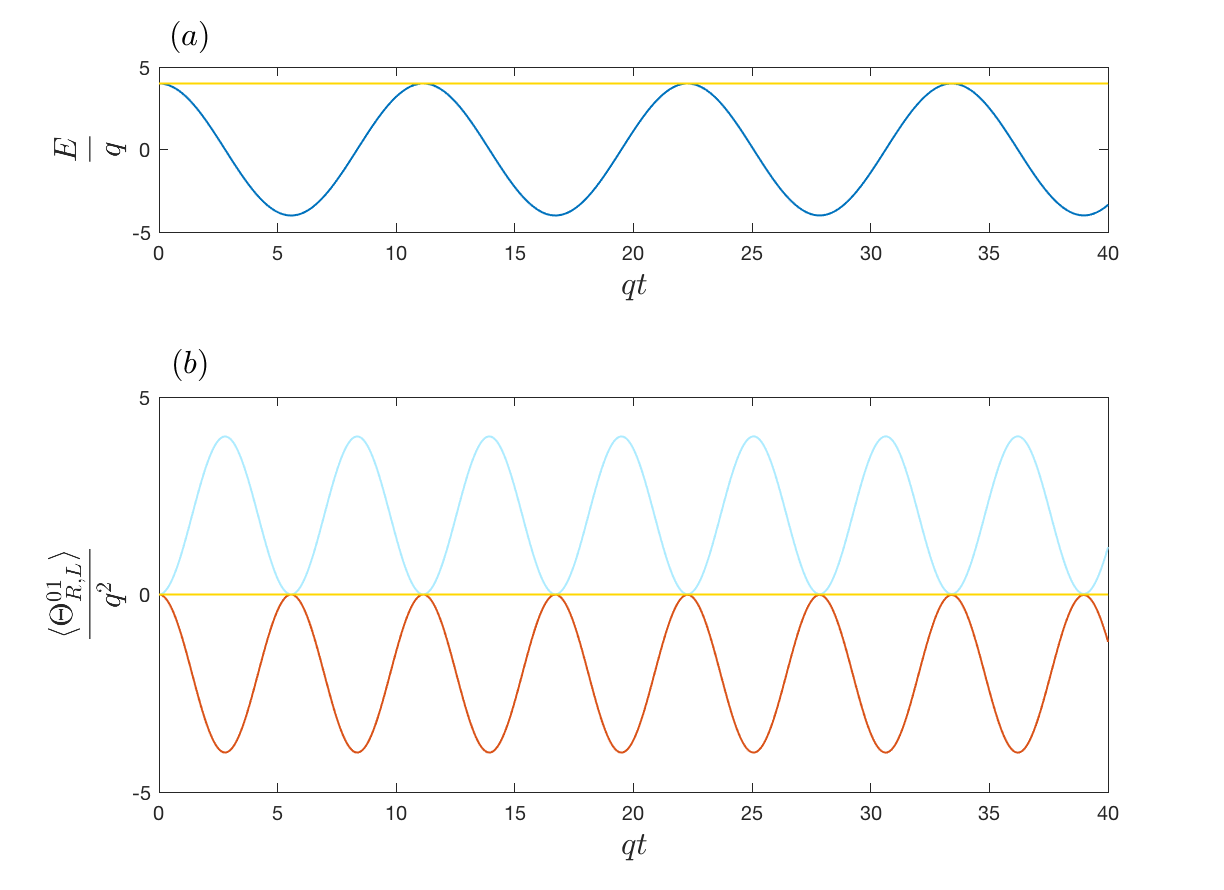} 
\end{tabular}
\end{center}
\caption{\small{Solution for the electric field (a) and the chiral projections of $\langle \Theta^{01}_R \rangle_{ren} $ (light blue line) and $\langle \Theta^{01}_L \rangle_{ren} $ (dark orange line) (b) for $m=0$. We have chosen $E_0=4q$ as the initial condition for the electric field. The initial state for the matter field is the vacuum.  The solution for the classical limit is also plotted (yellow line).}}
\label{F1}
\end{figure}

In the massless limit, the system can be solved analytically, finding harmonic oscillations with frequency $\frac{|q|}{\sqrt{\pi} }$. In Fig. \ref{F1}(a) we show the solution for the electric field. 
 
  It is very illuminating  to see the time evolution of the fluxes $\langle \Theta^{01}_{R, L} \rangle_{ren} $, since they represent the created chiral momentum in the massless case. 
  As we can see in  Fig. \ref{F1}.(b), for each set of massless right-handed  fermions/antifermions created with total momentum $P_R>0$, there is a set of massless left-handed antifermions/fermions with momentum $P_L=-P_R$.  The required energy to create particles is extracted from the electric field, generating a continuous energy exchange between the electric and the fermionic fields. Particles can also be destroyed returning  energy to the electric field.

In the massive case $m\neq 0$, the backreaction equations also induce electric  oscillations, which can be regarded as perturbations of the oscillations at $m=0$.  \\

\section{Conclusions} 

In this brief article, we have pointed out that quantized chiral  fields in two dimensions coupled to a homogeneous time-varying electric field break the classical conservation of the canonical stress-energy tensor $\partial_\mu \langle T^{\mu 1}_{R,L}\rangle_{ren} = {\mp} \frac{q^2 A\dot A}{2\pi}$. This quantum anomaly  has not been stressed in the previous literature. This result can be reexpressed in terms of the symmetric stress-energy tensor of the left- or right-moving sectors of the Dirac field $\partial_\mu \langle \Theta^{\mu 1}_{R,L}\rangle_{ren} = {\pm} \frac{q^2 A\dot A}{2\pi}$. Furthermore, our results have    a direct physical interpretation in terms of particle creation and in a way compatible with the axial anomaly. \\

{\it Acknowledgments.--}  We thank I. Agullo, A. del Rio, and A. Ferreiro for very useful discussions. This work was supported by  Grants.  No.\  FIS2017-84440-C2-1-P, No. \ FIS2017-91161-EXP, No. \ SEJI/2017/042 (Generalitat Valenciana) and 
No. \  SEV-2014-0398. P. B. is supported by a Ph.D. fellowship, Grant No.  FPU17/03712. S. P.  is supported by a Ph.D. fellowship, Grant No. FPU16/05287. 

\end{document}